\documentclass[a4paper]{article}

\usepackage[english]{babel}
\usepackage[utf8x]{inputenc}
\usepackage[T1]{fontenc}
\usepackage{bbm}
\usepackage[table,xcdraw]{xcolor}
\usepackage{booktabs}
\usepackage{url}
\usepackage{authblk}

\usepackage[a4paper,top=3cm,bottom=2cm,left=3cm,right=3cm,marginparwidth=1.75cm]{geometry}

\usepackage{amsmath}
\usepackage{graphicx}

\usepackage[colorlinks=true, allcolors=blue]{hyperref}

\title{A generalization of the symmetrical and optimal probability-to-possibility transformations}

\author[1]{Esteve del Acebo}
\affil[1]{VICOROB Institute, Universitat de Girona}
\author[2]{Yousef Alizadeh-Q}
\author[2]{Sayyed  Ali Hossayni}
\affil[2]{Data Mining Lab, School of Computer Engineering.  \authorcr Iran University of Science and Technology}

\begin{document}
\maketitle

\begin{abstract}

Possibility and probability theories are alternative and complementary ways to deal with uncertainty, which has motivated over the last years an interest for the study of ways to transform probability distributions into possibility distributions and conversely. This paper studies the advantages and shortcomings of two well-known discrete probability to possibility transformations: the optimal transformation and the symmetrical transformation, and presents a novel parametric family of probability to possibility transformations which generalizes them and alleviate their shortcomings, showing a big potential for practical application. The paper also introduces a novel fuzzy measure of specificity for probability distributions based on the concept of fuzzy subsethood and presents a empirical validation of the generalized transformation usefulness applying it to the text authorship attribution problem. 
\end{abstract}

\section{Introduction}

Possibility and probability theories are alternative ways to deal with uncertainty \cite{zadeh78}, \cite{Dubois:2007}. They are in no way unrelated (actually, a degree of possibility can be viewed as an upper probability bound  \cite{Dubois1993}) and are complementary in the sense that both can be useful under different circumstances. This has motivated over the last years the study of ways to obtain possibility distributions from probability distributions and conversely. These probability to possibility transformation are, citing Sudkamp \cite{SUDKAMP199273} purely mechanical transformations of probabilistic support to possibilistic support and viceversa. That is, a conversion of the measure of support of one theory into that of the other that is independent of the problem domain.

Several such transformations have been defined (see \cite{Ous00} for a detailed account), two of the most well-known being the optimal transformation and the symmetrical transformation proposed by D. Dubois \textit{et al.} \cite{Dubois82,Dubois1993}. Both transformations have their drawbacks, which we will discuss: the discontinuity in the case of the optimal transformation and the low specificity in the case of the symmetrical transformation. This paper presents a new parametric family of probability to possibility transformations which generalizes the optimal transformation and the symmetrical transformation and can contribute to alleviate their shortcomings. The paper is organized as follows: we first present the two transformations, examine their properties and expose its possible deficiencies. In the next section we show how the two transformations can be seen as particular cases of a parametric family of transformations and study their properties, introducing a novel fuzzy measure of specificity for possibility distributions. Section 4 deals with converse possibility to probability transformations, giving a formulation of the converse of the optimal transformation and also giving the system of equations to be solved in order to obtain the converse of the generalized transformation. Section 5 presents a empirical validation of the generalized transformation usefulness applying it to the text authorship attribution problem and, finally, the last section
contains several concluding remarks.

\section{Probability to possibility transformations}

As stated in the introduction, a number of probability to possibility and possibility to probability transformations, both in the continuous and the discrete cases, have been proposed. In this section, we will study two well-known discrete probability to possibility transformations proposed by D. Dubois \emph{et al.}: the so-called symmetric \cite{Dubois82} and optimal \cite{Dubois1993} transformations. We will examine their properties and discuss their advantages and drawbacks.

\subsection{Possibility distributions and possibility measures}Let $W=\{w_1,w_2,...,w_n\}$ be the set of possible values taken by a discrete random variable $X$, let $p:W \rightarrow [0..1]$ be the probability distribution of $X$ and $P:2^W \rightarrow [0..1]$ be the probability measure induced by $p$. A possibility distribution for $X$ is a function $\pi:W \rightarrow [0..1]$ and can be seen as a fuzzy set over $W$. The possibility measure induced by $\pi$ is defined, for all $A \subseteq W$ as $\Pi(A)=\max_{w_i \in A}(\pi(w_i))$

\subsection{The optimal and the symmetrical transformations}
Following Zadeh \cite{zadeh78}, Dubois and Prade propose several desirable properties for a probability to possibility transformation \cite{Dubois82}:
\begin{itemize}
\item \textit{Consistence.} An event must be possible prior to being probable. Degrees of possibility cannot be less than degrees of probability. $\forall A  \subseteq W \ \Pi(A) \geq P(A)$
\item \textit{Order preservation.} Probabilities and possibilities must be equally ordered. $\forall w_i,w_j \in W \ \pi(w_i) > \pi(w_j) \iff p(w_i) > p(w_j)$
\end{itemize}
In \cite{Dubois:1983}, Dubois \textit{et al.} define a transformation with these properties. It is known as the symmetrical transformation $\pi_{S}$, and is defined as:
\begin{equation}
\label{symmetricaltransformation}
\pi_{S}(w_i)=\sum_{w_j \in W}min(p(w_i),p(w_j))
\end{equation}

Another desirable property for a probability to possibility  transformation is \textit{maximal specificity,} in the sense that the possibility distribution should preserve as much information from the probability distribution as possible. We will say that a possibility distribution $\pi_1$ is more specific than another possibility distribution $\pi_2$  iif $ \forall w_i \in W \ \pi_{1}(w_i) \leq \pi_{2}(w_i).$ (That is, iif $\pi_1 \subset \pi_2$, denoting $\subset$ fuzzy inclusion.) The maximally specific probability to possibility transformation satisfying the properties of consistence and order preservation was presented by D. Dubois \textit{et al.} in \cite{Dubois1993}; it is known as the \textit{optimal transformation} (proof can be found in \cite{Dubois1993}, \cite{Delgado87}) and is defined as: 
\begin{equation}
\label{optimaltransformation1}
\pi_{O}(w_i)=\sum_{w_j \in W/p(w_j)\leq p(w_i)}p(w_j)
\end{equation}
or, equivalently:
\begin{equation}
\label{optimaltransformation2}
\pi_{O}(w_i)=\sum_{w_j \in W}p(w_j)\cdot \mathbbm{1}_{\leq w_i}(w_j)
\end{equation}
where $\mathbbm{1}_{\leq w_i}$ is the indicator function equal to one if $w_j \leq w_i$ and zero otherwise.

Both transformations have their advantages and shortcomings. The symmetrical transformation is intuitive and continuous (more on this later) but has as a shortcoming its low specificity, it preserves less information from the underlying probability distribution than the optimal transformation. The optimal transformation, on the other hand, is maximally specific, but is discontinuous in a way that makes it counter-intuitive in some cases. For example, let $p=[0.5,0.5]$ represent the probability distribution function of a binary random variable $X$. The associated possibility distributions using the symmetrical and the optimal transformations would be $\pi_S = \pi_O= [1,1]$. If we change the probabilities to $p=[0.501,0.499]$  we will have $\pi_S=[1,0.998]$, but $\pi_O=[1,0.499]$. So, in the case of the optimal transformation, arbitrarily small changes in the probability distribution function can produce large changes in the possibility distribution function. This is not the case with the symmetrical transformation.

\begin{figure}
\centering
\includegraphics[width=0.85\textwidth]{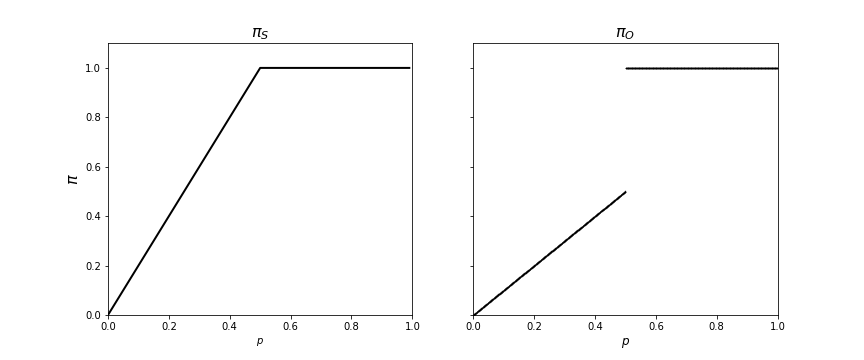}
\caption{\label{fig:pradestransformations}Plot of $\pi(w_1)$ against $p(w_1)$ for a binary random variable $X$ taking values in $W=\{w_1,w_2\}$ with probabilities $p(w_1)=p$ and $p(w_2)=1-p$. To the left, the symmetrical transformation $\pi_S$, to the right, the optimal transformation $\pi_O$.}
\end{figure}

A further example which makes this discontinuity evident graphically can be seen in figure \ref{fig:pradestransformations}. It plots $\pi(w_1)$ against $p(w_1)$ for a binary random variable $X$ taking values in $W=\{w_1,w_2\}$ with probabilities $p(w_1)=p$ and $p(w_2)=1-p$. To the left, the plot of $\pi_S$ is continuous, to the right, the plot of $\pi_O$ shows a discontinuity at $p=0.5$

\section{A generalized transformation}
As we have seen in the previous section, both transformations have shortcomings: the symmetrical transformation is not specific enough and the discontinuity of the optimal transformation  is apt to produce counter-intuitive results. It could be desirable to have the possibility of trading part of the specificity of the later for the continuity of the former. Our proposal is to generalize both transformations by means of the following family of parametric transformations
\begin{equation} \label{eq:generalized}
\pi_G(w_i)= \sum_{w_j \in W}p(w_j) \cdot min \bigg(1,{\Big({\frac{p(w_i)}{p(w_j)}\Big)}^n\bigg)} 
\end{equation}

It is easy to see that $\pi_S$ and $\pi_O$ are particular cases of $\pi_G$. For $n=1$, clearly $\pi_G=\pi_S$. On the other hand, when $n$ tends to infinity, $min\big(1,\big( \frac{p(w_i)}{p(w_j)} \big)^n \big)$ tends to the indicator function $\mathbbm{1}_{\leq w_i}$ and, consequently, $\pi_G$ tends to $\pi_O$. 
\begin{figure}
\centering
\includegraphics[width=0.5\textwidth]{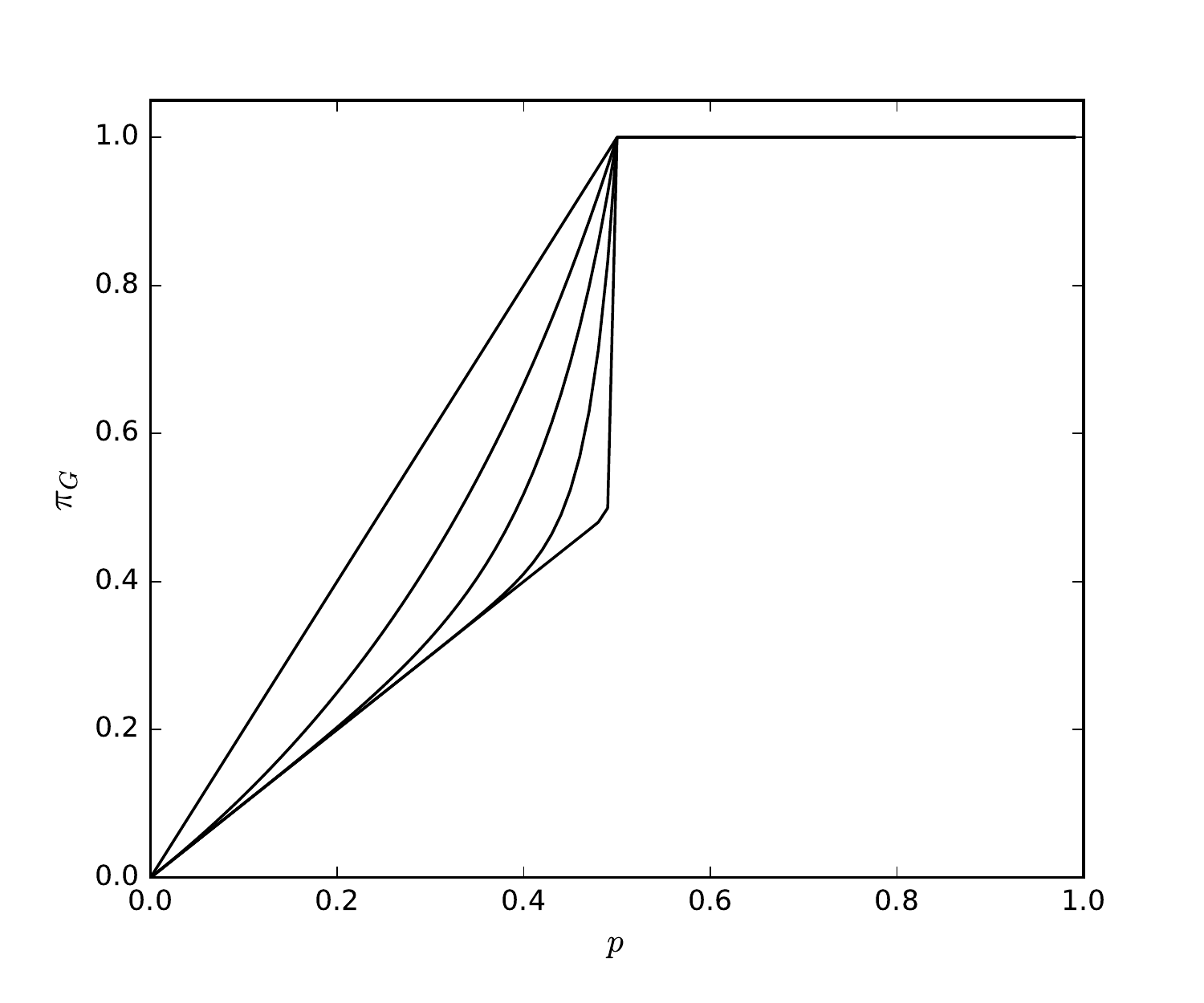}
\caption{\label{fig:generalized}Plots of $\pi_G(w_1)$ against $p(w_1)$ for a binary random variable taking values in $W=\{w_1,w_2\}$ with probabilities $p(w_1)=p$ and $p(w_2)=1-p$ and for several values of the parameter $n$. From left to right, $n=1,2,4,10,100$}
\end{figure}

\begin{figure}
\centering
\includegraphics[width=0.75\textwidth]{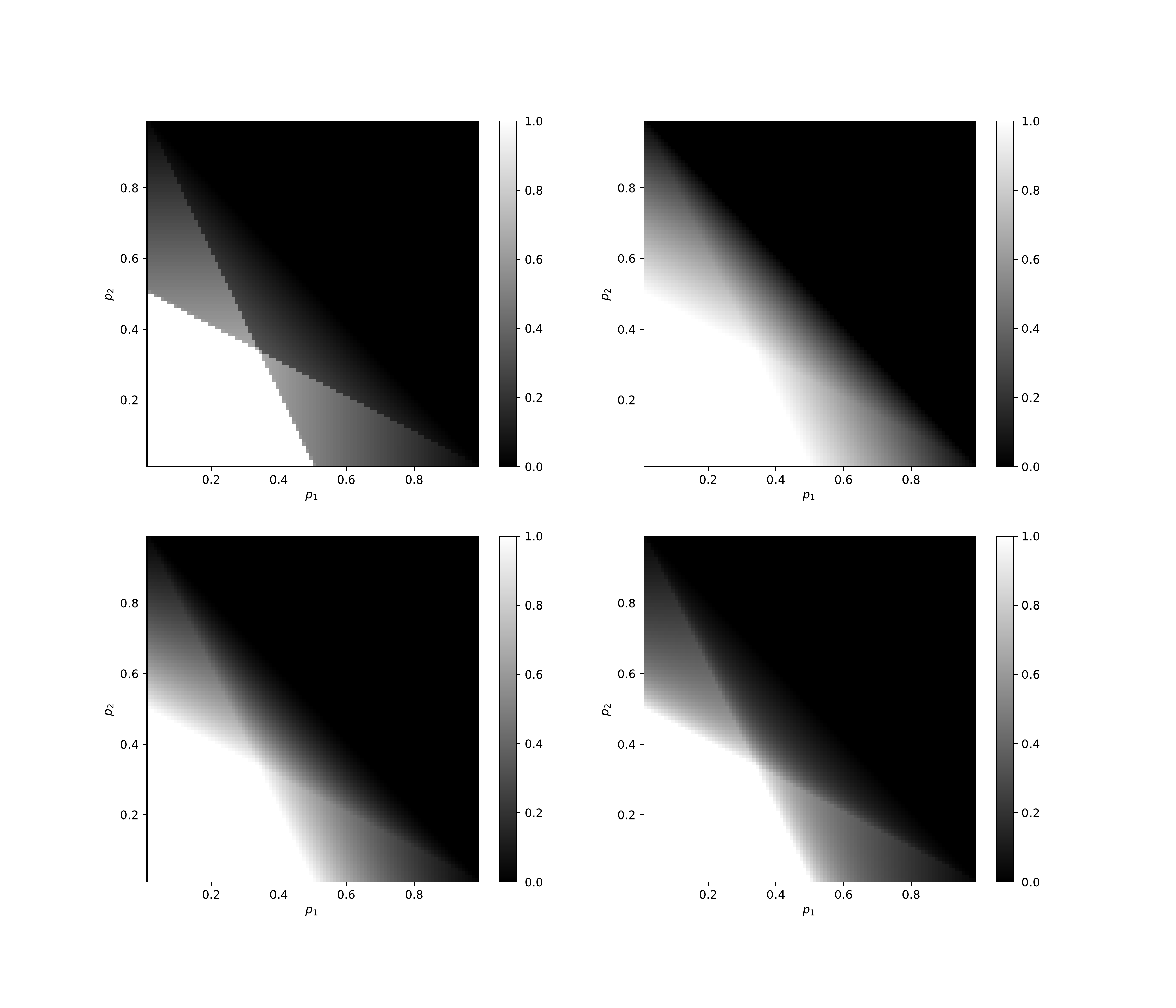}
\caption{\label{fig:colormaps}color maps showing $\pi_G(w_3)$ against $p_1=p(w_1)$ and $p_2=p(w_2)$ for a ternary random variable taking values in $W=\{w_1,w_2,w_3\}$ for different values of $n$. In the top row, to the left, $n=100$ ($\pi_G \approx \pi_O$). To the right, $n=1$, $(\pi_G=\pi_S$. In the bottom row,  to the right $n=2$, to the left $n=5$.}
\end{figure}

In Fig. \ref{fig:generalized} we can see plots of $\pi_G(w_1)$ against $p(w_1)$ for a binary random variable taking values in $W=\{w_1,w_2\}$ with probabilities $p(w_1)=p$ and $p(w_2)=1-p$ and for several values of the parameter $n$. From left to right, $n=1,2,4,10,100$. It can be seen the increase of specificity without lose of continuity. Similarly, in Figure \ref{fig:colormaps} we can see color maps showing $\pi_G(w_3)$ against $p_1=p(w_1)$ and $p_2=p(w_2)$ for a ternary random variable taking values in $W=\{w_1,w_2,w_3\}$ for different values of $n$. Only the regions where $p_1 + p_2 \leq 1$ (that is, under the main diagonal) are meaningful. In the top row, to he left, for $n=100$,  $\pi_G \approx \pi_O$. It is easy to see the discontinuity lines of $\pi_O$. To the right, for $n=1$, $\pi_G=\pi_S$ and no discontinuities exist, but there is low specificity. In the bottom row, $\pi_G$ for two intermediate values of $n$: left, $n=2$, right, $n=5$.
\subsection{Properties of the generalized transformation}
It is easy to prove that the generalized transformation has the properties of consistence and order preservation for any value $n > 0$. Consistence can be proved from the properties of the optimal transformation: it holds that $min\big(1,\big( \frac{p(w_i)}{p(w_j)} \big)^n \big)\geq \mathbbm{1}_{\leq w_i}(w_j) $ for all $n$. So, for all $A \subseteq W $, we will have $ \Pi_G(A)= \max_{w_i \in A}\pi_G(w_i) \geq \Pi_O(A) \geq P(A)$. In order to prove order preservation, we can observe in Eq. \ref{eq:generalized} that, for any pair of probabilities $p(w_1)$ and $p(w_2)$:
\begin{itemize}
\item $p(w_1) = p(w_2) \implies \pi(w_1) = \pi(w_2)$ 
\item If $p(w_1) > p(w_2)$, every term in the sum for the computation of $\pi_G(w_1)$ is greater or equal than the corresponding term for $\pi_G(w_2)$ and, particularly, the term corresponding to $w_j=w_1$ is strictly greater in the former. So $p(w_1) > p(w_2) \implies \pi(w_1) > \pi(w_2)$
\end{itemize}
\subsection{Specificity of the generalized transformation}

In this section, we will try to measure and compare the specificity of the generalized transformation for different values of the parameter $n$ and different probability distributions. We have said that a possibility distribution $\pi_1$ is more specific than another possibility distribution $\pi_2$ if $\pi_1(w_i) <\pi_2(w_i)$ for all $w_i \in W$. This is the same than saying that a possibility distribution $\pi_1$ is more specific than another possibility distribution $\pi_2$ if $\pi_1 \subset \pi_2$, considering $\pi_1$ and $\pi_2$ fuzzy sets and denoting $\subset$ fuzzy inclusion. Following this idea, it makes sense to make use of a fuzzy subsethood relationship to extend this definition of specificity to the fuzzy domain and talk about the degree to which $\pi_1$ is more specific than $\pi_2$. We will use the fuzzy subsethood relationship proposed by Kosko \cite{kosko90}: given two fuzzy sets $A$ and $B$ over the same universal set $U$, we define the degree to which $A$ is contained in $B$ as:  
\begin{equation}
S(A,B)=\frac{M(A\cap B)}{M(A)}
\end{equation}
where $\cap$ denotes fuzzy intersection and $M(X)$ denotes the cardinality or measure of the fuzzy set $X$, defined as $M(X) = \sum_{u_i \in U}m_X(u_i)$. Making use of this definition and choosing the minimum operator to implement fuzzy intersection, we define, given a probability distribution $p$, the specificity of the possibility distribution $\pi_T$ obtained by applying a probability to possibility transformation $T$ to $p$ as the degree to which it is included into the maximally specific possibility distribution $\pi_O$ (the one obtained by applying the optimal transformation to $p$):
\begin{equation}
specificity(\pi_T)=S(\pi_T,\pi_O)=\frac{\sum_{w_i\in W}\min (\pi_T(w_i),\pi_O(w_i))}{\sum_{w_i\in W} \pi_T(w_i)}
\end{equation}
The results of a series of experiments can be seen in table \ref{exp_results}. Each row shows, for a given value of the exponent $n$ in equation \ref{eq:generalized}, the mean and the standard deviation of the specificity of 100 possibility distributions resulting from the generalized transformation of a sample of 100 probability distributions. Each probability distribution $p_i$ is obtained sampling 250000 times a discrete random variable $V$ taking values over $W = \{w_1,w_2,...,w_{1000}\}$ and then assigning to $p_i(w_i)$ the proportion of occurrences of $w_i$ in the sample.  To the left, the results when $V$ is distributed uniformly over $W$, that is,  $p(w_i) = \frac{1}{1000}$. To the left, the results when the probability distribution of $V$ follows a power law, where $p(w_i)$ is proportional to $\frac{1}{i^\alpha}$. This distribution is known as Zipf's law or discrete Pareto distribution, and is known for its usefulness to model many types of data studied in the physical and social sciences, from frequency of words in natural language corpuses to population ranks of cities in various countries, corporation sizes, income rankings, ranks of number of people watching the same TV channel and so on \cite{wiki:zipf_law},

The results show, as expected, how the specificity of the transformed possibility distribution increases with the value of parameter $n$. It is interesting to observe the difference in the growth speed depending upon the underlying probability distribution. The specificity of the possibility distributions obtained from the power law probability distribution increases much more rapidly than the specificity of the possibility distributions obtained from uniform probability distribution.

The experiment has been run several times, with different values of the number of samples and different cardinality of $W$, the results being similar to those reported in the table.
\begin{table}[]
\centering
\caption{Mean and standard deviation of the specificity of a set of possibility distributions for different values of the parameter $n$ in equation \ref{eq:generalized}. To the left, results when the probability  distributions are obtained sampling a random variable distributed uniformly. To the right, results when the probability distributions are obtained sampling a random variable distributed following Zipf's law  }
\label{exp_results}
\begin{tabular}{@{}ccccc@{}}
\toprule
           & \multicolumn{2}{c}{\textbf{Uniform}} & \multicolumn{2}{c}{\textbf{Zipf's}} \\ \midrule
%\rowcolor[HTML]{FFFFFF} 
\textbf{n} & \textbf{specificity}  & \textbf{SD}  & \textbf{specificity}  & \textbf{SD} \\
1          & 0.5098                & 0.0002       & 0.5050                & 0.0001      \\
2          & 0.5272                & 0.0002       & 0.6744                & 0.0005      \\
3          & 0.5439                & 0.0005       & 0.7596                & 0.0007      \\
5          & 0.5751                & 0.0011       & 0.8443                & 0.0007      \\
10         & 0.6416                & 0.0021       & 0.9201                & 0.0005      \\
100        & 0.9318                & 0.0014       & 0.9956                & 0.0001      \\ \bottomrule
\end{tabular}
\end{table}
\section{Converse transformations}
The symmetrical probability to possibility transformation defined in eq. \ref{symmetricaltransformation} has a well-known \cite{Dubois:1983} corresponding converse possibility to probability transformation given by:
\begin{equation}
p_S(w_i)=\sum_{i \leq j \leq n}\frac{\pi_S(w_j)-\pi_S(w_{j+1})}{j}
\end{equation}
considering, without loss of generality, that possibility (and, consequently, probability) values are ordered decreasingly and taking  $\pi_S(w_{n+1})=0$.

In the case of the optimal probability to possibility transformation defined in eqs. \ref{optimaltransformation1} and \ref{optimaltransformation2}, if all the $\pi(w_i)$ are different, the converse possibility to probability transformation is as simple as follows:
\begin{equation}
p(w_i)=\pi_O(w_i)-\pi_O(w_{i+1})
\end{equation}
also considering $\pi_O(w_{n+1})=0$, but is very easy to see that it does not work when possibility values repeat \footnote{In fact, this is the converse of the transformation $\pi(w_i)=\sum_{j \geq i}p(w_j)$, a variation of the optimal transformation which replaces the \textit{order preservation} condition with the \textit{weak order preservation} condition given by: $\forall w_i,w_j \in W  p(w_i) > p(w_j) \Rightarrow \pi(w_i) > \pi(w_j)$. It is the most specific consistent transformation \cite{Delgado87}, but has the drawbacks that it is not unique and that no possibility values can repeat, even if the corresponding probability values are equal. That is, for all $i \neq j \ \pi(w_i) \neq \pi(w_j)$ }. 
As far as we know, no published formulation of the converse of the optimal transformation taking into account the possibility of duplicated values exists. We give it as:
\begin{equation}
p_O(w_i)=\frac{\pi_O(w_i)-\max (\pi_O(w_j)/\pi_O(w_j)<\pi_O(w_i))}{reps(\pi_O(w_i))}
\end{equation}
Where $reps(\pi_O(w_i))$ is the number of repetitions of the value $\pi_O(w_i)$ in the possibility distribution and also considering $\pi_O(w_{n+1})=0$. In other words, given a decreasingly ordered list of possibility values, the probability of $w_i$, $p_O(w_i)$ equals to its possibility $\pi_O(w_i)$, minus the next value in the possibilities list different from $\pi_O(w_i)$, if such value exist, and divided by the number of times the value $\pi_O(w_i)$ appears in the list.

The general case is, however, much more involved. Suppose the probabilities $p(w_1) ... p(w_M)$ ordered decreasingly and let $p_i$ and $\pi_i$ denote $p(w_i)$ and $\pi(w_i)$ respectively. The generalized transformation can be written as:
\begin{equation}
\pi_i=\sum_{j=1}^{i-1}p_j \Big(\frac{p_i}{p_j}\Big)^n + \sum_{j=i}^{M}p_j = p_i^{n}\sum_{j=1}^{i-1}p_j^{n-1}+ \sum_{j=i}^{M}p_j 
\end{equation}
This expression defines, supposing the $\pi_i$'s known, a system of $M$ equations in $M$ variables (the $p_i$'s) which must be solved under the restriction $p_1 \geq p_2 \geq ... \geq p_M > 0$. This is a hard problem without a general closed-form algebraic solution, to the best of our knowledge, which has to be solved with specialized mathematical software. Moreover, the existence of a solution is not clear to be guaranteed for every possibility distribution. 

\section{Empirical validation}
In a previous work in forensic document analysis \cite{DBLP:journals/corr/abs-1902-02176, Sayyed2018}, the authors showed how the fusion of two author characteristics (the stylome or specific style of writing and the author's handwritten signature) can be used in the text authorship attribution problem in order to improve attribution accuracy for several linear classificators. To this end, two biometric algorithms were combined, a fuzzy signature recognition algorithm due to Kudlacik and Porwik \cite{Kudlacik2014} and a probabilistic authorship attribution algorithm from Sidorov \emph{et al.} \cite{Sidorov}. In order to combine the results of the algorithms, it was necessary to homogenize them by transforming the probabilistic output of the authorship attribution algorithm to possibility values. This was done using both the optimal and symmetrical probability to possibility transformations, with the result that the symmetrical transformation provided more accurate results. In this section we will prove empirically the usefulness of the generalized transformation presented in this paper by repeating the experiments done in \cite{DBLP:journals/corr/abs-1902-02176, Sayyed2018} using the generalized transformation with different values of the parameter $n$.
\begin{figure}
\centering
\includegraphics[width=0.85\textwidth]{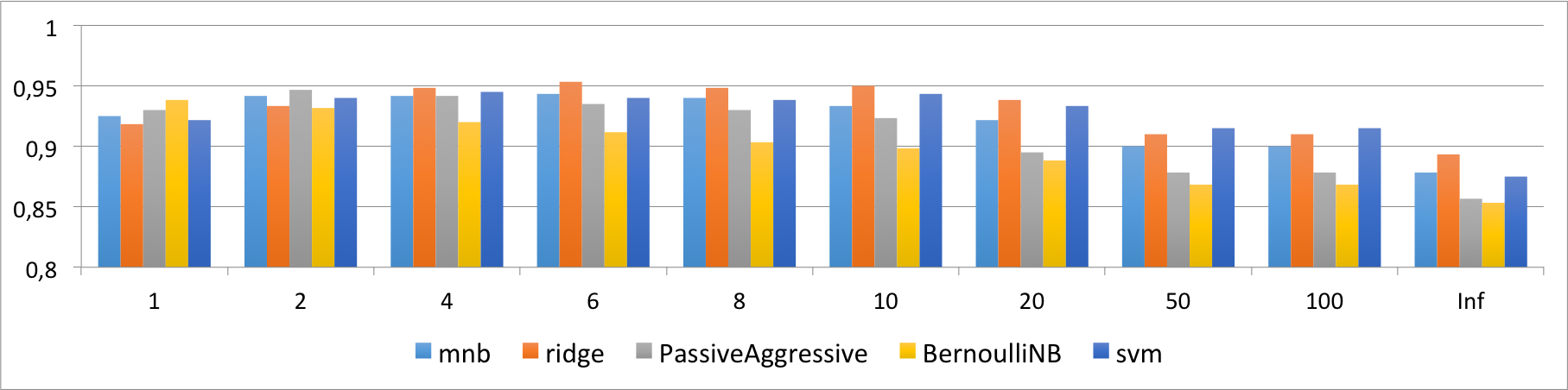}
\caption{\label{fig:possibility} Accuracy of the author attribution method proposed in \cite{DBLP:journals/corr/abs-1902-02176} for different values of the parameter $n$. There are five results for each value of $n$, corresponding to the use of five different linear-time classifiers inside the syntactic n-gram-based authorship attribution algorithm. From left to right: MNB stands for Multinomial Naïve Bayes, SVM stands for Support Vector Machine, Ridge stands for Ridge-Regression classification  PA stands for Passive-Aggressive classification and BNB stands for Bernoulli Naïve Bayes classification. Results for $n=1$ correspond to the symmetrical transformation $\pi_S$, results for $n=inf$ correspond to the optimal transformation $\pi_O$.}
\end{figure}

The mail database used in the experiments contains a total of 800 mails from 40 authors (20 mails each). For each mail, the text body and the scanned author's handwritten signature are available. 5 mails from each author are used for training and the remaining 15 are used for testing. For each test mail we compute both the fuzzy membership of its signature to the set of training signatures of each author using Kudlacik and Porwik algorithm \cite{Kudlacik2014} and the probabilities of authorship of its text body by each author using Sidorov \emph{et al.} algorithm \cite{Sidorov} and five different linear-time classifiers (the algorithm uses one classifier itself). Thereafter, the obtained probabilities are transformed to possibilities using the generalized transformation and, after fusion, the set (i.e. the author) with the maximum membership is chosen as the genuine author.

We can see the outcome of the experiments in \ref{fig:possibility}. The $x$ axis represent the value of the generalized transformation parameter $n$ and the $y$ axis represent the accuracy of the corresponding classifier. There are five results for each value of $n$, corresponding to the use of five different linear-time classifiers inside the syntactic n-gram-based authorship attribution algorithm. From left to right: MNB stands for Multinomial Naïve Bayes, SVM stands for Support Vector Machine, Ridge stands for Ridge-Regression classification \cite{dobriban2018}, PA stands for Passive-Aggressive classification \cite{Matsu}, and BNB stands for Bernoulli Naïve Bayes classification. Results for $n=1$ correspond to the symmetrical transformation $\pi_S$ and results for $n=inf$ correspond to the optimal transformation $\pi_O$)

As can be seen in the figure, when the attribution algorithm uses Multinomial Naïve-Bayes or Ridge-Regression classification, the best results correspond to $n = 6$; when it uses Passive-Aggressive classification the best results correspond to $n=2$ and when it uses Bernoulli Bayes they correspond to $n=1$, that is, the symmetrical transformation $\pi_{S}$ . More important, the best accuracy overall is obtained with the attribution algorithm using Ridge Regression classification with values of $n$ ranging from 4 to 10. We believe that this results provide strong empirical evidence of the possible usefulness of the generalized transformation. 

\section{Conclusions}
This paper has presented a novel parametric family of discrete probability to possibility transformations which generalize two well know transformations proposed by D. Dubois \textit{et al.} \cite{Dubois82, Dubois1993}, the symmetrical transformation and the optimal transformation, making possible to combine to different degrees their advantages by increasing the specificity of the symmetric transformation without losing continuity and avoiding, in this way, possible artifacts caused by the lack of continuity of the optimal transformation. This gives the presented generalized transformation a big potential for practical application.
We have also proved that the generalised transformation has the properties of consistence and order preservation for positive values of the exponent $p$, and devised a fuzzy measure for possibility distributions specificity based on a fuzzy subsethood relationship. Finally, we have given empirical evidence of the usefulness of the generalized transformation by comparing it with the symmetrical and optimal transformations in the context of an author attribution problem.

It remains as further work to do to establish to what extent this generalized transformation represents a real improvement over the existing ones, analyzing the importance of the value of the parameter $n$ and studying how to determine which values are best suited for a given application. It would also be interesting to determine which numerical or algebraic methods could be suitable for the calculation of the dual possibility to probability transformation for different values of the exponent $n$.  

\section*{Acknowledgements}
This work has been partially granted by AfterDigital Consultants: Digitalización del consultor digital, RTC-2017-6370-7, CIEN Service Chain (Nuevas tecnologías basadas en blockchain para gestión de la identidad, confiabilidad y trazabilidad de las transacciones de bienes y servicios) and the Grup de Recerca Consolidat ref. 2017 SGR 1648.

\bibliographystyle{alpha}
\bibliography{main}

\end{document}